\documentclass[twocolumn,aps,10pt,showkeys,showpacs,preprintnumbers,prd,superscriptaddress,nofootinbib]{revtex4-1}
\usepackage{graphicx}	
\usepackage{amssymb}
\usepackage{dcolumn}
\usepackage{mathtools}
\usepackage{amsmath}
\usepackage{xcolor}
\usepackage{color}
\usepackage[normalem]{ulem}
\usepackage{theorem}
\usepackage{subfigure, rotating, bm, array}
\usepackage[pagebackref=false, colorlinks=true]{hyperref}
\hypersetup{linkcolor=magenta, citecolor=magenta,urlcolor=magenta}

\begin{document}

\title{Signatures of a Schwarzschild-like Black Hole Immersed in Dark Matter Halo}

\author{Akshat Pathrikar}
\email{akshatpathrikar014@gmail.com}
\affiliation{International Centre for Space and Cosmology, Ahmedabad University, Ahmedabad 380009, Gujarat, India}


\begin{abstract}
\noindent
Astrophysical black holes are often surrounded by dark matter, which can influence their dynamics and observational signatures. In this work, we study a Schwarzschild-like black hole immersed in a Dehnen-type 
$(1,4,2)$ dark matter halo and analyse scalar, electromagnetic, and gravitational perturbations in this spacetime. We compute quasinormal modes (QNMs) using the Wentzel-Kramers-Brillouin (WKB) approximation method with Padé approximants, investigate particle motion and photon trajectories, and use black hole shadow observations 
to place constraints on the halo parameters. We further examine the greybody factors associated with Hawking radiation for different perturbation spins. This combined analysis aims to understand how dark matter environments may affect black hole oscillations, radiation properties, and the corresponding observational signatures.

\vspace{.6cm}
$\boldsymbol{Key words}$ : Quasinormal modes, Black Holes, Dark Matter, Shadow, Greybody Factors
\end{abstract}

\maketitle

\section{Introduction}
\label{sec:intro}

Black holes (BHs) stand among the most fascinating and profound predictions of Einstein’s field equations in General Relativity (GR). Over the past century, GR has been subjected to an extraordinary range of observational and experimental tests, spanning both weak-field and strong-field regimes, and remarkably, it has passed each of them with overwhelming success. Classic tests in the weak-field regime include the gravitational lensing of distant galaxies \cite{Bartelmann:2010fz}, the anomalous precession of Mercury’s perihelion \cite{Park:2017zgd}, and the gravitational redshift of light in a gravitational field \cite{Hohensee:2011wt}. In the strong-field regime, highly compact astrophysical systems such as binary pulsars have provided precise confirmations of GR through their orbital decay via gravitational radiation.

In recent years, technological breakthroughs have enabled even more direct and striking tests of GR. The detection of gravitational waves by LIGO–Virgo \cite{LIGOScientific:2016aoc} has opened an entirely new observational window into the dynamical, strong-gravity regimes, providing evidence of black hole mergers and allowing precise measurements of spacetime geometry. Furthermore, the Event Horizon Telescope (EHT) has provided the first direct observational image of a black hole shadow \cite{EventHorizonTelescope:2019dse}, offering unprecedented insight into the near-horizon region where gravity is extreme. 

Despite the remarkable success of classical General Relativity, several fundamental issues remain unresolved within the framework of the theory. One of the most significant open problems is the nature of Dark Matter (DM), an unseen component that dominates the mass content of the universe but does not interact electromagnetically. GR accurately describes how dark matter gravitates, yet it offers no explanation for its microscopic origin or composition. Moreover, observations reveal that black holes are not isolated objects in the universe. Instead, they are typically surrounded by various forms of matter, such as galactic halos or dark matter halos that extend over large scales. In particular, supermassive black holes (SMBHs), which reside at the centers of most galaxies, are known to power Active Galactic Nuclei (AGN) \cite{Rees:1984si, Kormendy:1995er}, some of the most energetic phenomena in the cosmos. The interaction between the central black hole and the surrounding matter significantly influences galaxy evolution, feedback processes, and accretion dynamics \cite{Iocco:2015xga, Bertone:2018krk}.

The first compelling indication for the existence of dark matter emerged from the observation that the outer regions of spiral and elliptical galaxies rotate much faster than what would be expected if only visible matter were present \cite{Persic:1995ru}. This discrepancy implied the presence of an additional, unseen mass component. Since then, a wide range of astrophysical observations has reinforced this conclusion, showing that dark matter constitutes the majority of a galaxy’s total mass, while ordinary baryonic matter contributes only a small fraction \cite{Sofue:2000jx}. During galaxy formation, dark matter is believed to have played a central role in gravitational collapse and structure formation, eventually redistributing into extended halos surrounding the visible components of galaxies \cite{Primack:1997av, Silk:2006df}.

Observations further indicate that most massive galaxies host a supermassive black hole at their center, and this black hole typically lies within the surrounding dark matter halo \cite{Valluri:2002xs, EventHorizonTelescope:2019ggy}. The coexistence of these two components suggests that dark matter may influence the dynamics of the galactic core and potentially leave detectable signatures in strong-gravity environments. On cosmological scales, measurements of the cosmic microwave background show that dark matter makes up about 27\% of the total energy content of the universe, while only about 5\% corresponds to ordinary matter and the remainder is dark energy \cite{Planck:2018vyg}.

Despite its dominant gravitational role, the fundamental nature of dark matter remains unknown, as it is not explained within the Standard Model of particle physics. This has motivated the proposal of numerous candidates, including weakly interacting massive particles, axions, and sterile neutrinos \cite{Boehm:2003hm, Bertone:2004pz, Feng:2009mn, Schumann:2019eaa}. Since dark matter interacts very weakly with ordinary matter, one of the most effective ways to study its properties is through its gravitational influence on astrophysical systems. In particular, dark matter may accumulate around supermassive black holes, affecting processes such as gravitational wave emission in extreme and intermediate mass-ratio inspirals and altering the surrounding spacetime structure \cite{Babak:2017tow, Brown:2006pj}. Additionally, phenomena such as galactic rotation curves and colliding galaxy clusters (e.g., the Bullet Cluster) provide strong observational support for the existence of dark matter as a distinct gravitational component \cite{ Clowe:2006eq}. 

Therefore, studying such astrophysical black hole systems embedded in matter is essential not only for understanding black hole physics, but also for probing the nature of dark matter and testing gravity in complex, realistic environments. These investigations provide a crucial bridge between fundamental physics, and astrophysical observations.\\

Investigations of Dehnen-type dark matter halos have explored the interaction between black holes (BHs) and their surrounding DM environments from various perspectives \cite{Dehnen:1993uh, Gohain:2024eer, Pantig:2022whj, Al-Badawi:2024asn}. For example, some studies have examined how the inner slope of the halo density profile influences the survival of low star-formation efficiency star clusters after rapid gas expulsion \cite{Shukirgaliyev:2021}. Other works have considered star clusters modelled with Plummer and Dehnen profiles, focusing on how different initial cusp slopes affect their evolution. In addition, a BH embedded in a Dehnen-type DM halo has been proposed as a model for ultra-faint dwarf galaxies \cite{Pantig:2022whj}. More recently, new BH solutions surrounded by Dehnen-type DM halos have been developed, often employing a Schwarzschild BH within a DM background. Within these setups, analyses have been carried out on the thermodynamic properties and null geodesics of the effective BH--DM halo system \cite{ Xamidov:2025hrj}, followed by studies that constrain the parameters of the DM halo.
Furthermore, the influence of the DM halo on QNMs, the photon sphere radius, and the BH shadow has been investigated in these frameworks \cite{Al-Badawi:2025njy, Liang:2025vux, Uktamov:2025lsq}, and the resulting gravitational waveforms from periodic orbits have also been explored \cite{Alloqulov:2025ucf}.
In this work, we focus on a recently proposed solution to the Einstein field equations \cite{Uktamov:2025lwb}, which describes a Schwarzschild-like black hole immersed in a dark matter halo characterized by a Dehnen-type density profile $(1,4,2)$. The construction of the model, together with the analysis of curvature invariants and the verification of the associated energy conditions, is presented in \cite{Uktamov:2025lwb}.

The paper is organized as follows. In Section (\ref{sec:intro}), we provide the necessary background and motivation. In Section (\ref{sec:GR2}), we introduce the Black Hole--DM halo solution and discuss its main properties. Section (\ref{sec:GR3}) is devoted to the study of perturbations and the corresponding physical implications. In Section (\ref{sec:GR4}), we introduce the WKB method and present all the QNM frequencies from Table (\ref{Tab1}) to (\ref{Tab12}) and analyse our results. In Section (\ref{sec:GR5}) we work our the particle motion and obtain constraints on the parameters of the metric and in Section (\ref{sec:GR6}) we present the corresponding Greybody factors. Finally, we conclude our findings in section (\ref{sec:GR7}) with possible future directions.

\section{Schwarzschild-like black hole spacetime with dark matter halo}
\label{sec:GR2}

The goal of this work is to consider a Schwarzschild-like black hole spacetime surrounded by a dark matter halo characterized by a Dehnen-type density profile. Our approach begins with a background spacetime sourced by the DM distribution, and the BH--DM solution is then obtained by solving Einstein’s field equations. In this framework, the Schwarzschild BH is embedded in a halo whose mass distribution follows the Dehnen profile, determined by the DM density (see details in \cite{Mo:2010gfe}). 

We first describe the spacetime geometry of the DM halo and subsequently incorporate the Schwarzschild BH geometry. To proceed, we introduce the general density profile in order to derive the corresponding Dehnen-type DM halo mass distribution, which is defined as
\begin{equation}
\rho(r) = \rho_s \left( \frac{r}{r_s} \right)^{-\gamma}
\left[ \left( \frac{r}{r_s} \right)^{\alpha} + 1 \right]^{\frac{\gamma-\beta}{\alpha}},
\label{eq:density}
\end{equation}
where $\rho_s$ and $r_s$ denote the characteristic density and scale radius of the DM halo, respectively. In addition to these parameters, the constants $\alpha$, $\beta$, and $\gamma$ characterize the slope of the density profile; in particular, $\gamma$ takes values in the range $0 \leq \gamma \leq 3$. In this work, we adopt the specific Dehnen-type profile $(\alpha, \beta, \gamma) = (1, 4, 2)$ for the DM halo.

Using Eq.~\eqref{eq:density}, the enclosed mass profile is given by
\begin{equation}
M_D(r) = \int_0^r 4\pi \rho(r_1) r_1^2 \, dr_1 = 4\pi \rho_s r_s^3 \left( 1 + \frac{r_s}{r} \right).
\label{eq:massprofile}
\end{equation}

The line element for the DM halo spacetime, written in terms of the redshift function $A(r)$ and the shape function $B(r)$, is
\begin{equation}
ds^2 = -A(r)\,dt^2 + \frac{dr^2}{B(r)} + r^2 d\Omega^2,
\label{eq:dm-metric}
\end{equation}
where $d\Omega^2 = d\theta^2 + \sin^2\theta\, d\phi^2$ denotes the solid angle in spherical coordinates.

An important feature of this setup is that $A(r)$ can be related to the tangential velocity of test particles moving in circular orbits within the halo:
\begin{equation}
v_D^2 = \frac{1}{r} \frac{d}{dr} \left( \log \sqrt{A(r)} \right) = \frac{M_D}{r},
\label{eq:velocity}
\end{equation}
which allows us to determine $A(r)$ as
\begin{equation}
\begin{aligned}
A(r) &= \left( 1 + \frac{r_s}{r} \right)^{-8\pi r_s^2 \rho_s} \\[4pt]
&\simeq 1 - \frac{2M_D(r)}{r_s} 
\left( 1 + \frac{r_s}{r} \right)
\log \left( 1 + \frac{r_s}{r} \right).
\label{eq:A}
\end{aligned}
\end{equation}

For the halo metric \eqref{eq:dm-metric}, the Einstein field equations take the form \cite{ }:
\begin{equation}
R_{\mu\nu} - \tfrac{1}{2} g_{\mu\nu} R = 8\pi T_{\mu\nu}^{(D)},
\label{eq:EFE-DM}
\end{equation}
where the energy-momentum tensor of the Dehnen halo is
\begin{equation}
T^\nu_{\ \mu (D)} = \mathrm{diag}[-\rho(r),\, P_r(r),\, P_t(r),\, P_t(r)] .
\end{equation}

To include the BH, we write the combined BH--DM halo metric as
\begin{equation}
ds^2 = -\left[A(r) + F_1(r)\right] dt^2
+ \frac{dr^2}{B(r) + F_2(r)} 
+ r^2 d\Omega^2,
\label{eq:combined}
\end{equation}
for which the Einstein field equations become
\begin{equation}
R_{\mu\nu} - \tfrac{1}{2} g_{\mu\nu} R
= 8\pi \left[ T_{\mu\nu}^{(D)} + T_{\mu\nu}^{(\text{BH})} \right],
\label{eq:EFE-total}
\end{equation}
where $T_{\mu\nu}^{(\text{BH})}$ is the energy-momentum tensor associated with the BH geometry.

By combining Eqs.~\eqref{eq:EFE-DM} and \eqref{eq:EFE-total} with the metrics \eqref{eq:dm-metric} and \eqref{eq:combined}, we obtain the following relations:
\begin{eqnarray}\label{eq.comp.Einsteinfield1}
\Big[B(r)&+&F_2(r)\Big]\Big[\frac{1}{r}\frac{B'(r)+F_2'(r)}{B(r)+F_r(r)}+\frac{1}{r^2}\Big]\nonumber\\&=& B(r)\Big[\frac{1}{r}\frac{B'(r)}{B(r)}+\frac{1}{r^2}\Big],
\end{eqnarray}
\begin{eqnarray}\label{eq.comp.Einsteinfield2}
\Big[B(r)&+&F_2(r)\Big]\Big[\frac{1}{r}\frac{A'(r)+F_1'(r)}{A(r)+F_1(r)}+\frac{1}{r^2}\Big]\nonumber\\&=& B(r)\Big[\frac{1}{r}\frac{A'(r)}{A(r)}+\frac{1}{r^2}\Big]\,
\end{eqnarray}

These equations lead to the space-time metric, including the DM halo, which can be written as follows  
\begin{eqnarray} \label{m14}
ds^{2}&=& -\exp \left[ \int \frac{B(r)}{B(r)-\frac{2M}{r}}\left( \frac{1}{r}+
\frac{A^{\prime }(r)}{A(r)}\right)dr \right]dt^{2} \nonumber\\&-& A(r) dt^{2}+\frac{dr^{2}}{B(r)-\frac{2M}{r} }+r^{2}\left(
d\theta ^{2}+\sin ^{2}\theta d\phi ^{2}\right)\, ,\nonumber\\
\end{eqnarray}

In the absence of the DM halo, one recovers $A(r)=B(r)=1$, and the integral yields the standard Schwarzschild solution $(1 - 2M/r)$. As a result, solving Eqs.~\eqref{eq.comp.Einsteinfield1} and \eqref{eq.comp.Einsteinfield2} gives
\begin{eqnarray}\label{eq.F}
    F_1(r)&=& exp\Big[\int \left(\frac{B(r)}{B(r)+F_2(r)}\Big(\frac{1}{r}+\frac{A'(r)}{A(r)}\Big)-\frac{1}{r}\right)dr\Big]\nonumber\\&&-A(r)\, ,\\
    F_2(r)&=&-\frac{2M}{r}\, ,
\end{eqnarray}
\label{eq:F2}
where the prime denotes differentiation with respect to $r$.

For simplicity, we shall assume \( A(r) = B(r) \), the field equations yield a Schwarzschild-like spacetime modified by the presence of a dark DM halo characterized by a Dehnen-type density profile with parameters \((1, 4, 2)\). The exact analytical form of the metric obtained from the field equations~(\ref{eq:EFE-DM}) is given by  
\begin{equation}
ds^{2} = -f(r)\,dt^{2} + \frac{dr^{2}}{f(r)} + r^{2}\left(d\theta^{2} + \sin^{2}\theta\, d\phi^{2}\right),
\label{eq:metric-dehnen}
\end{equation}
where
\begin{equation}
f(r) = 1 - \frac{2M}{r}
- \frac{8\pi \rho_s r_s^{2}} 
{\left(1 + \frac{r_s}{r}\right)}
\left(1 + \frac{r_s}{r}\right)
\log\!\left(1 + \frac{r_s}{r}\right).
\label{eq:fr-dehnen-corrected}
\end{equation}

\begin{figure}[h!]
    \centering
    \includegraphics[width=0.49\textwidth]{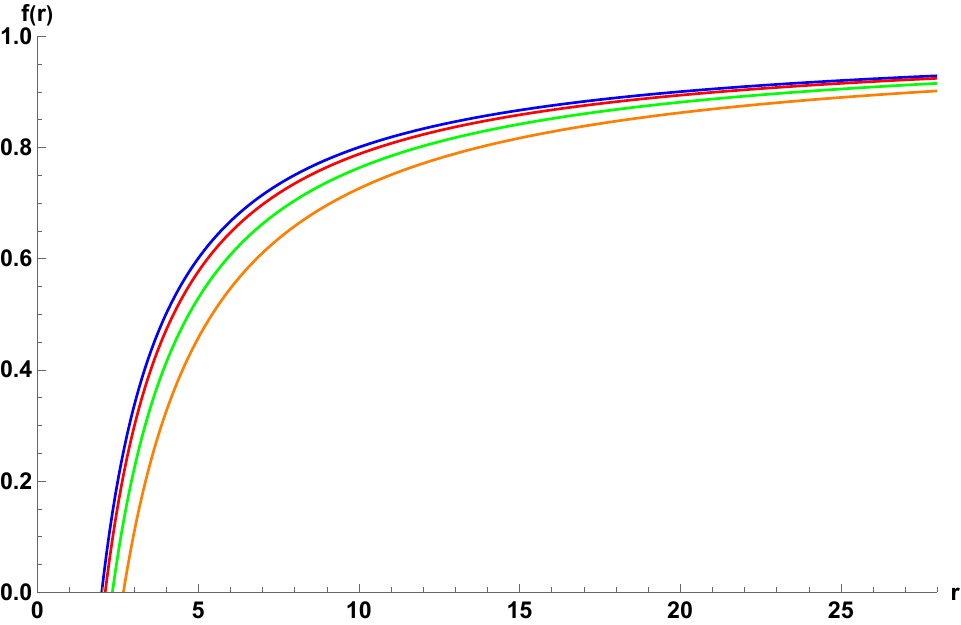}
    \caption{Variation of $f(r)$ with respect to the radial distance for various values of $\rho_{s} = 0.00$ (Blue), 0.01 (Red), 0.03 (Green) and 0.06 (Orange), we set M = 1 and $r_{s} = 0.8$.}

    \label{FIG1}
\end{figure}

\begin{figure}[h!]
    \centering
    \includegraphics[width=0.49\textwidth]{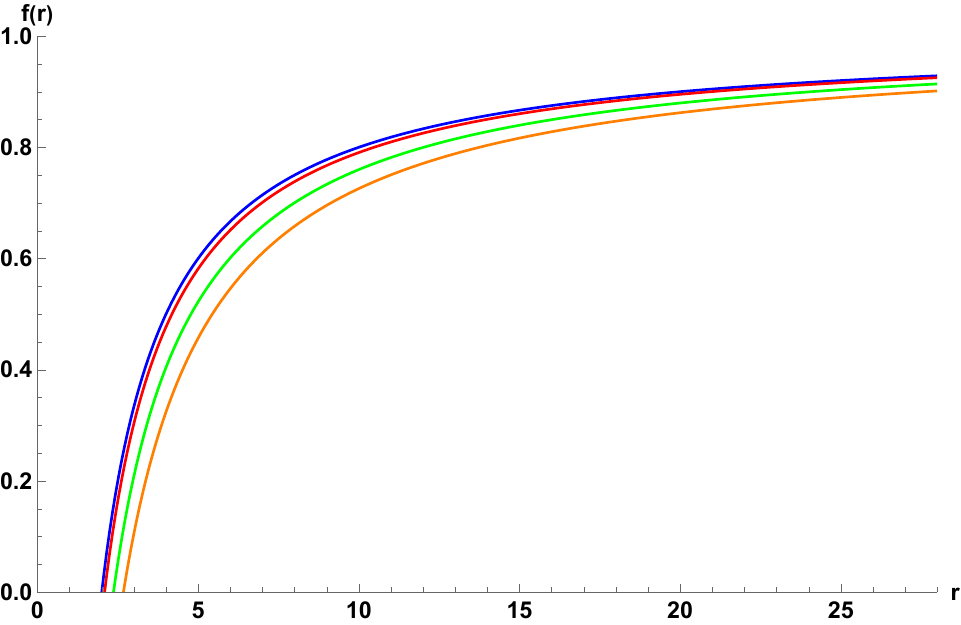}
    \caption{Variation of $f(r)$ with respect to the radial distance for various values of $r_{s} = 0.00$ (Blue), 0.04 (Red), 0.65 (Green) and 0.8 (Orange), we set M = 1 and $\rho_{s} = 0.06$}

    \label{FIG2}
\end{figure}

The radial behavior of \( f(r) \) is depicted in Figs.~\ref{FIG1} and \ref{FIG2} for different values of the halo parameters \( \rho_s \) and \( r_s \). As evident from the figure, the metric function asymptotically approaches unity at large radial distances, consistent with a flat spacetime limit. Moreover, increasing either \( \rho_s \) or \( r_s \) shifts the curve slightly toward larger \( r \), indicating that the dark matter halo enhances the overall gravitational potential.

\section{Wavelike equations and effective potentials}
\label{sec:GR3}


\subsection{Scalar and Electromagnetic Perturbations}

In this study, we analyze the behavior of various test fields propagating in the spacetime of a Schwarzschild-like black hole surrounded by the given DM halo. The test fields considered include scalar, electromagnetic, and gravitational perturbations. The general relativistic field equations describing the dynamics of the scalar field \((\Phi)\) and the electromagnetic potential \((A_\mu)\) are expressed as  
\begin{equation}
\frac{1}{\sqrt{-g}} \, \partial_\mu \!\left( \sqrt{-g} \, g^{\mu\nu} \partial_\nu \Phi \right) = 0,
\label{eq:scalar_field}
\end{equation}
\begin{equation}
\frac{1}{\sqrt{-g}} \, \partial_\mu \!\left( F_{\rho\sigma} g^{\rho\nu} g^{\sigma\mu} \sqrt{-g} \right) = 0,
\label{eq:em_field}
\end{equation}
where \( F_{\mu\nu} = \partial_\mu A_\nu - \partial_\nu A_\mu \) is the electromagnetic field strength tensor.  

After applying separation of variables to the background metric given in Eq.~(1), the above field equations reduce to a Schrödinger-like wave equation of the form~\cite{}:
\begin{equation}
\frac{d^2 \Psi}{dr_*^2} + \left( \omega^2 - V(r) \right) \Psi = 0,
\label{eq:wave_equation}
\end{equation}
where \( r_* \) denotes the tortoise coordinate, defined by  
\begin{equation}
dr_* = \frac{dr}{f(r)}.
\label{eq:tortoise}
\end{equation}
The effective potential for perturbations of spin \(s\) (with \(s = 0\) for scalar and \(s = 1\) for electromagnetic fields) is given by  
\begin{equation}
V(r) = f(r) \left[ \frac{\ell(\ell + 1)}{r^2} + (1 - s) \frac{1}{r} \frac{d^2 r}{dr_*^2} \right],
\label{eq:potential}
\end{equation}
where \(\ell = s,\, s+1,\, s+2,\, \ldots\) represent the multipole numbers.  

\begin{figure}[htbp]
\centering

\resizebox{\linewidth}{!}{%
\begin{tabular}{cc}
\includegraphics{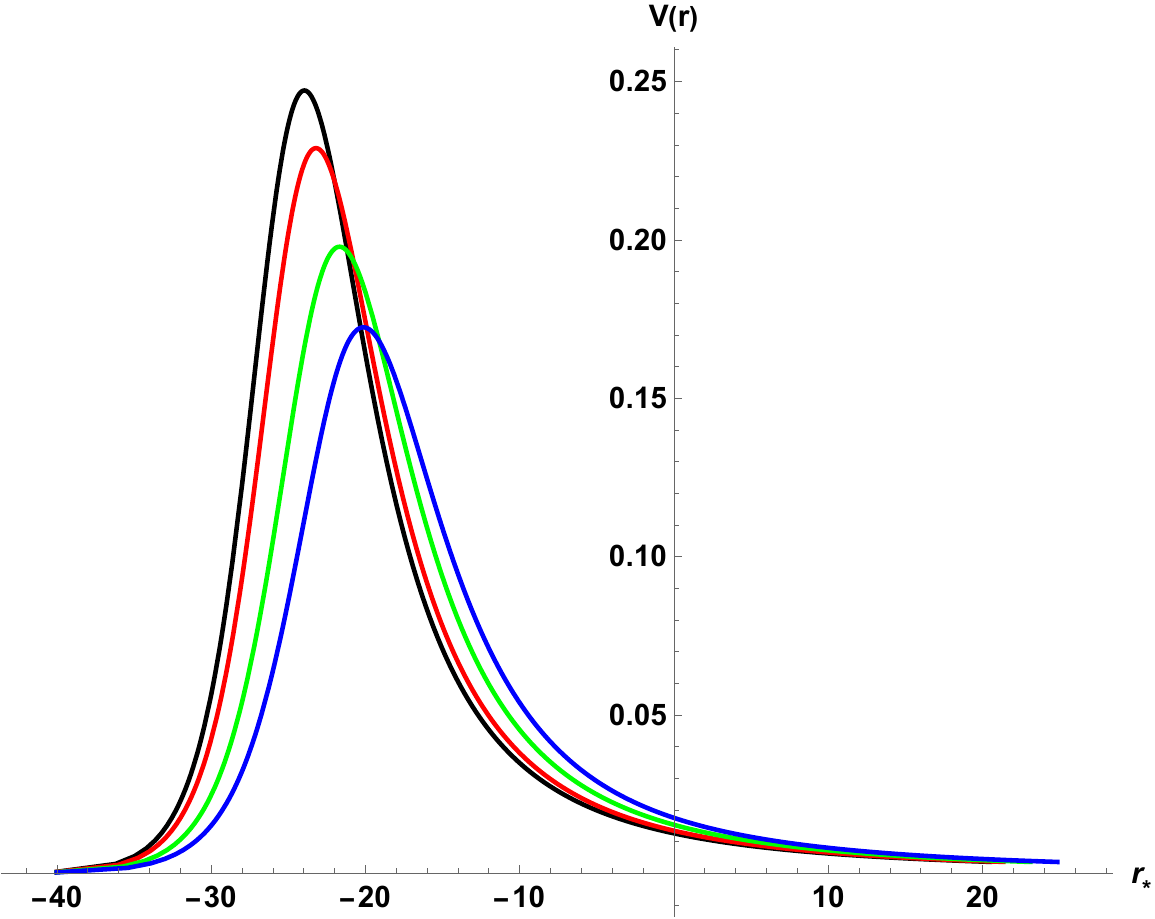} &
\includegraphics{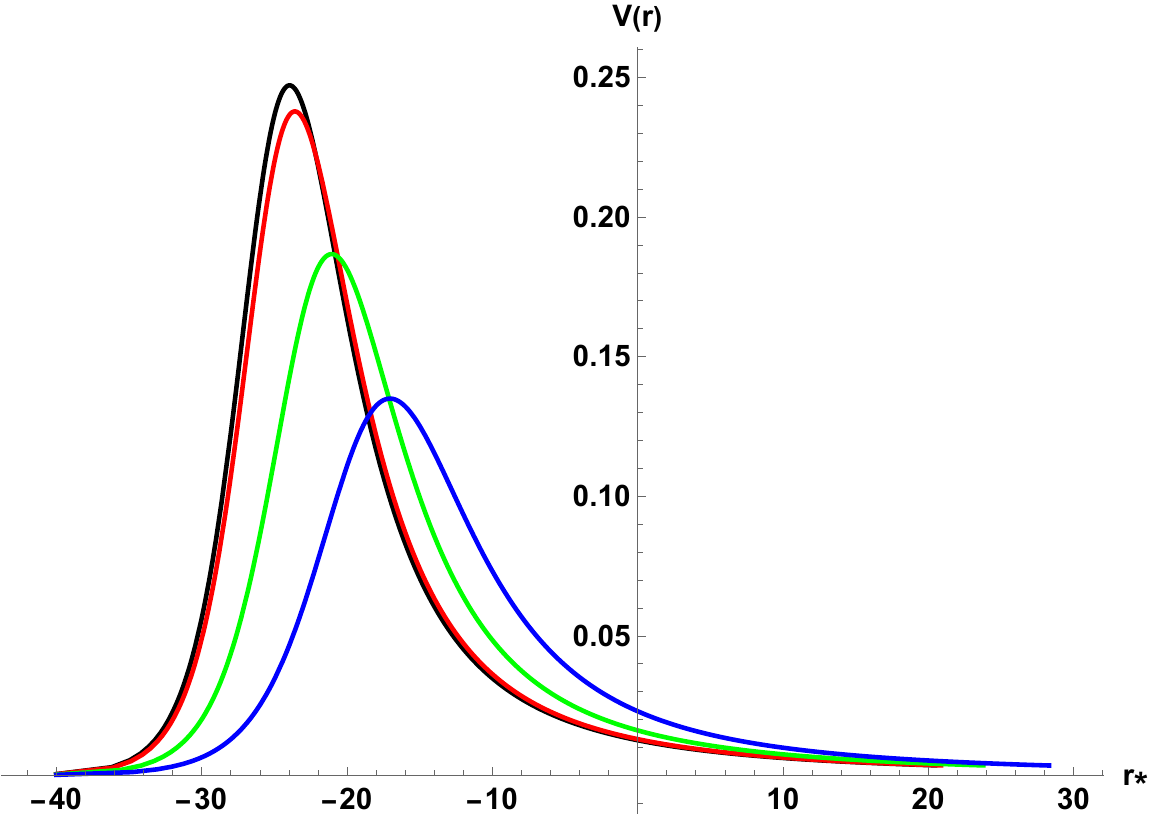} \\
\end{tabular}%
}
\caption*{Effective potential $V(r)$ versus tortoise coordinate $r_\ast$ for scalar field perturbations. 
Left: fixed $r_s = 0.7$ with varying $\rho_s$; Right: fixed $\rho_s = 0.06$ with varying $r_s$.}

\vspace{12pt}

\resizebox{\linewidth}{!}{%
\begin{tabular}{cc}
\includegraphics{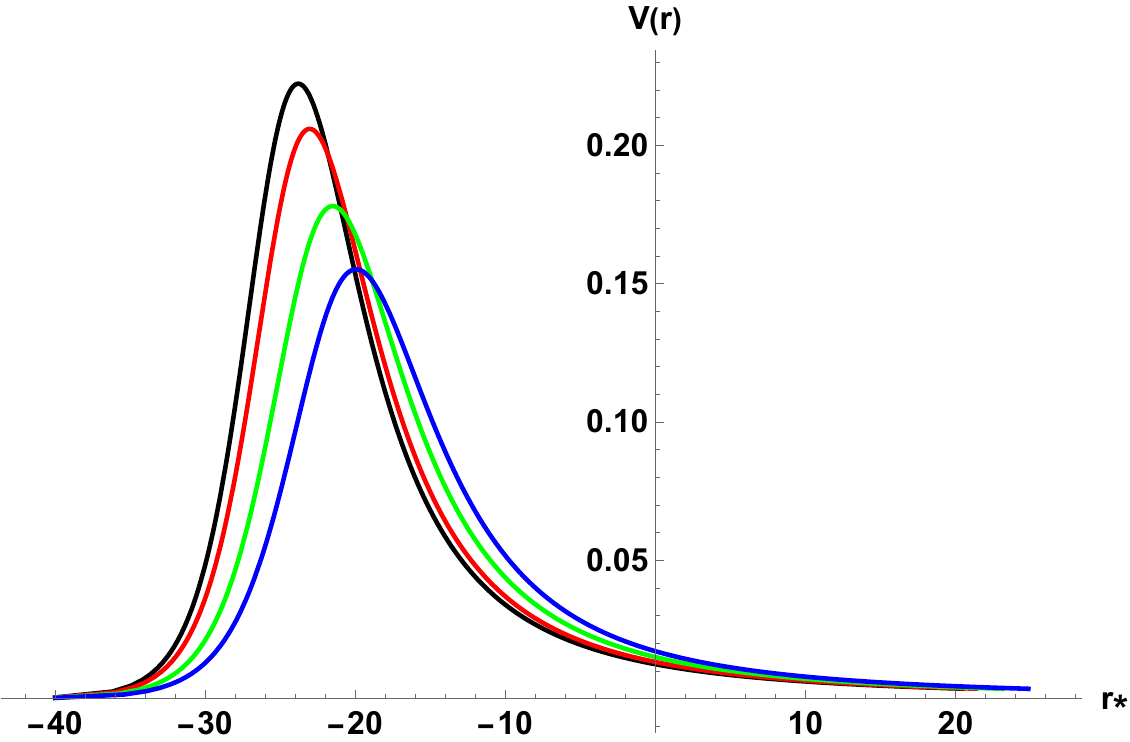} &
\includegraphics{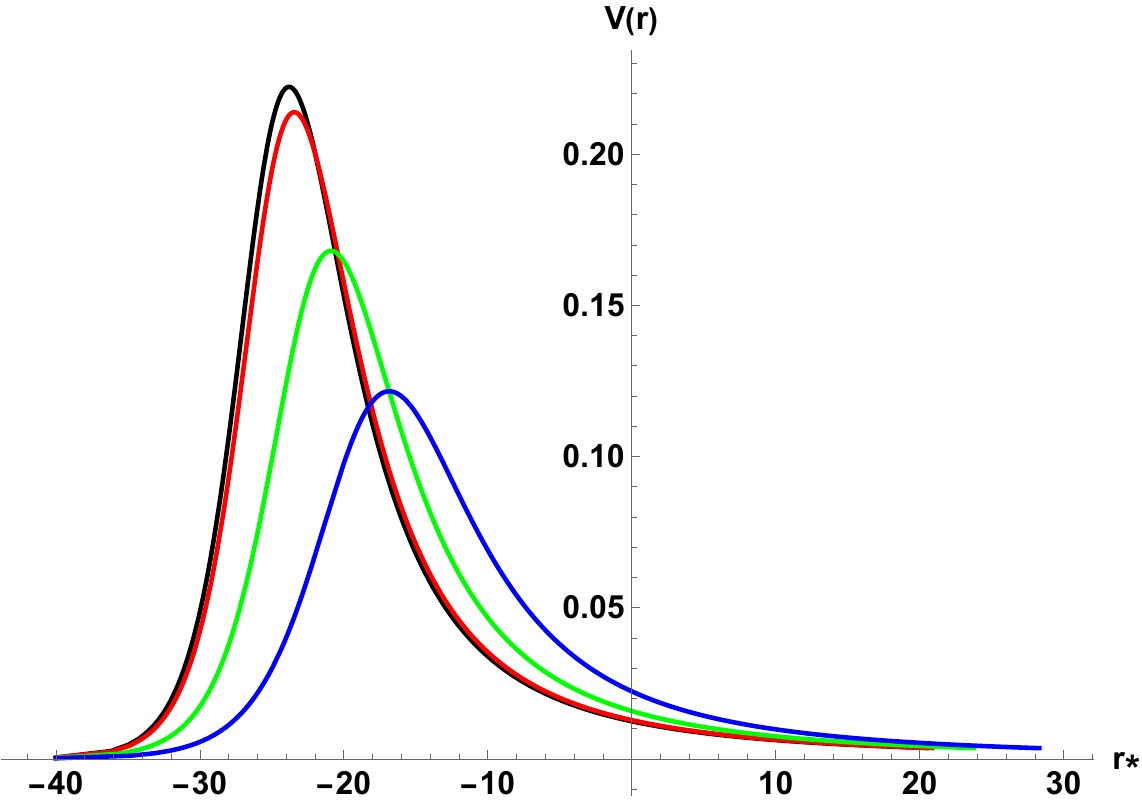} \\
\end{tabular}%
}
\caption*{Effective potential $V(r)$ versus tortoise coordinate $r_\ast$ for electromagnetic perturbations. 
Left: fixed $r_s = 0.7$ with varying $\rho_s$; Right: fixed $\rho_s = 0.06$ with varying $r_s$.}

\vspace{12pt}

\resizebox{\linewidth}{!}{%
\begin{tabular}{cc}
\includegraphics{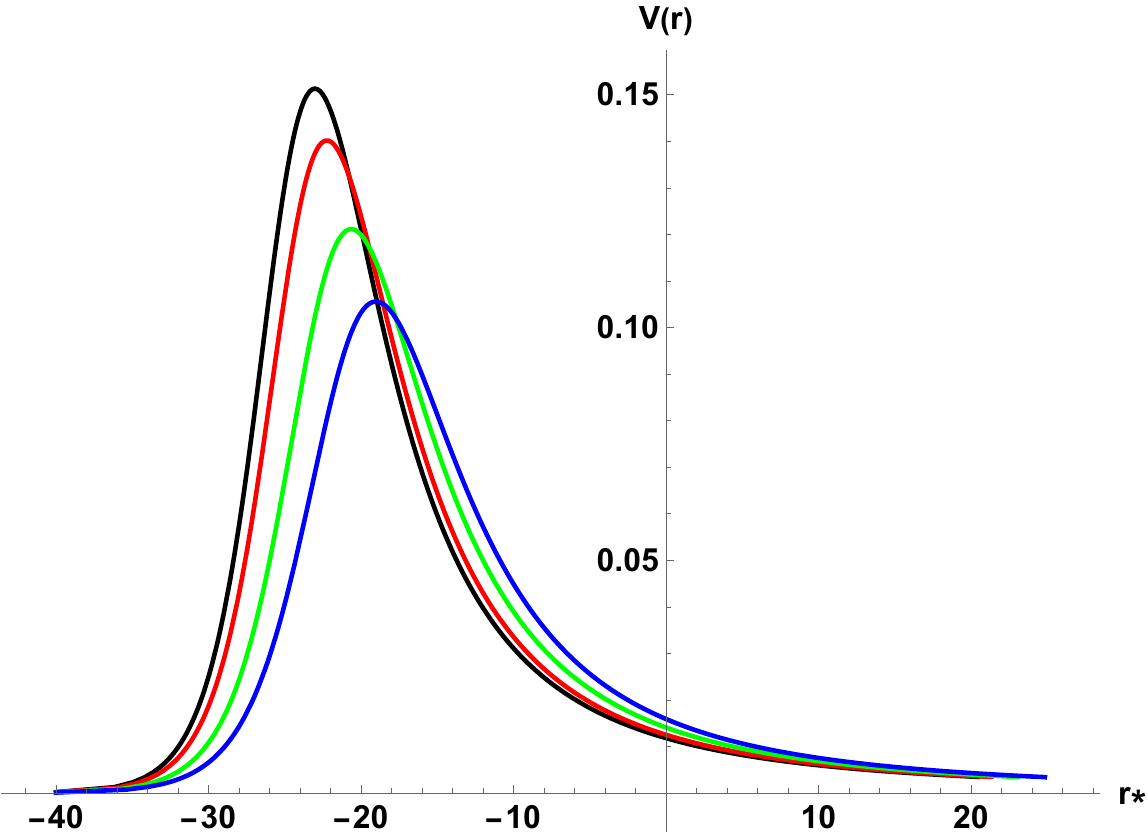} &
\includegraphics{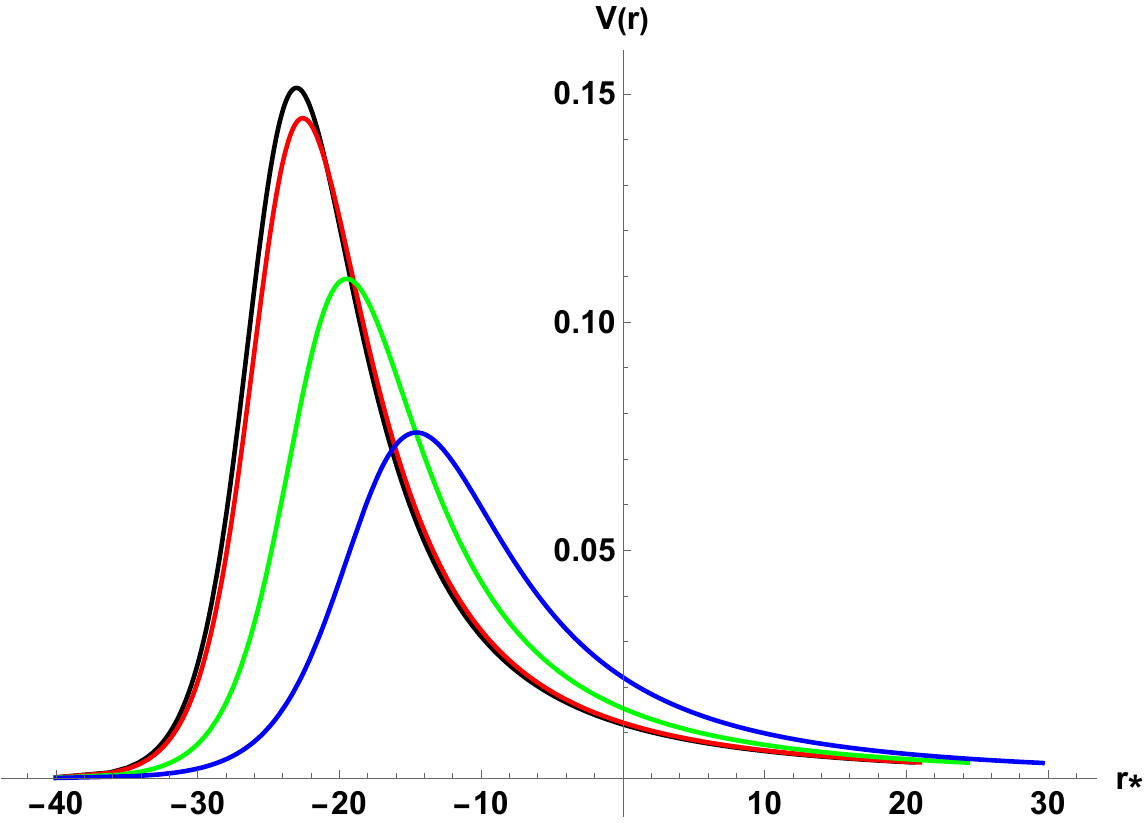} \\
\end{tabular}%
}
\caption*{Effective potential $V(r)$ versus tortoise coordinate $r_\ast$ for gravitational perturbations. 
Left: fixed $r_s = 0.7$ with varying $\rho_s$; Right: fixed $\rho_s = 0.07$ with varying $r_s$.}

\vspace{8pt}

\caption{
Combined effective potentials $V(r_*)$ for scalar, electromagnetic, and gravitational field perturbations 
in the Dehnen–$(1,4,2)$ dark matter halo ($M=1/2$). 
Each row corresponds to a different type of perturbation, while each column shows variation with the parameter $\xi$.}
\label{allpotentials}
\end{figure}

The corresponding effective potentials for the scalar, electromagnetic, and gravitational perturbations are illustrated in Fig.(\ref{allpotentials}). Since these potentials are positive definite, they ensure the stability of the respective field perturbations in the given background.

\subsection{Axial Gravitational Perturbations}

The axial perturbation of a black hole refers to small deviations in the spacetime geometry or matter distribution of the black hole that preserve axial symmetry. These perturbations can be effectively treated using perturbation theory. In this framework, the spacetime metric \( g_{\mu\nu} \) is written as a sum of the unperturbed background metric \( \bar{g}_{\mu\nu} \) and a small perturbation term \( h_{\mu\nu} \):
\begin{equation}
g_{\mu\nu} = \bar{g}_{\mu\nu} + h_{\mu\nu}.
\label{eq:metric_perturbation}
\end{equation}

Here, \( \bar{g}_{\mu\nu} \) represents the stable background geometry of the black hole when it is not significantly influenced by external factors. For axial perturbations, the perturbation term satisfies \( h_{\mu\nu} \ll \bar{g}_{\mu\nu} \).

The application of perturbation theory is not limited to the metric alone, it also affects the Christoffel symbols and the Ricci tensor. Thus, their perturbed forms can be written as
\begin{align}
\Gamma^{\lambda}_{\mu\nu} &= \bar{\Gamma}^{\lambda}_{\mu\nu} + \delta\Gamma^{\lambda}_{\mu\nu}, \label{eq:Gamma_perturbation}\\
R_{\mu\nu} &= \bar{R}_{\mu\nu} + \delta R_{\mu\nu}. \label{eq:Ricci_perturbation}
\end{align}
The perturbation terms \( \delta\Gamma^{\lambda}_{\mu\nu} \) and \( \delta R_{\mu\nu} \) are given by
\begin{align}
\delta\Gamma^{\lambda}_{\mu\nu} &= \frac{1}{2} \bar{g}^{\lambda\beta} 
\left( h_{\mu\beta;\nu} + h_{\nu\beta;\mu} - h_{\mu\nu;\beta} \right), \label{eq:deltaGamma}\\
\delta R_{\mu\nu} &= \delta\Gamma^{\lambda}_{\mu\lambda;\nu} - 
\delta\Gamma^{\lambda}_{\mu\nu;\lambda}. \label{eq:deltaR}
\end{align}

Here, \( \bar{\Gamma}^{\lambda}_{\mu\nu} \) and \( \bar{R}_{\mu\nu} \) are the Christoffel symbols and Ricci tensor corresponding to the background metric \( \bar{g}_{\mu\nu} \). The terms \( \delta\Gamma^{\lambda}_{\mu\nu} \) and \( \delta R_{\mu\nu} \) represent the influence of the metric perturbation \( h_{\mu\nu} \) on the spacetime connection and curvature. Since the perturbation of the background field makes no contribution relative to the background itself, it follows that \cite{Kobayashi:2012kh}
\begin{equation}
\delta R_{\mu\nu} = 0.
\label{eq:deltaR_zero}
\end{equation}

We now consider axial gravitational perturbations satisfying the Regge–Wheeler (RW) gauge~\cite{Kokkotas:1999bd}. This gauge exploits the spherical symmetry of the background and the properties of axial perturbations to impose constraints on the perturbed metric, leading to a solvable wave equation describing the perturbation behavior. Under this gauge, the perturbation term \( h_{\mu\nu} \) takes the form
\begin{equation}
h_{\mu\nu} =
\begin{pmatrix}
0 & 0 & 0 & h_0(t,r) \\
0 & 0 & 0 & h_1(t,r) \\
0 & 0 & 0 & 0 \\
h_0(t,r) & h_1(t,r) & 0 & 0
\end{pmatrix}
\sin\theta \, \partial_\theta P_\ell(\cos\theta),
\label{eq:axial_metric}
\end{equation}
where \( h_0(t,r) \) and \( h_1(t,r) \) are functions of the time \( t \) and the radial coordinate \( r \), describing the characteristics of the gravitational perturbation. \( P_\ell(\cos\theta) \) denotes the Legendre polynomial of order \( \ell \).

Substituting Eqs.~(\ref{eq:EFE-total}) and~(\ref{eq:Gamma_perturbation}) into Eq.~(\ref{eq:metric_perturbation}) yields
\begin{equation}
\begin{split}
\frac{\partial^2 \psi}{\partial t^2}
& - f \frac{\partial}{\partial r}
  \left( f \frac{\partial}{\partial r} (r\psi) \right)
 + \frac{2f^2}{r^2} \frac{\partial}{\partial r}(r\psi) \\
& + f\!\left[ \frac{\ell(\ell+1)}{r^2}
 - \frac{2f'}{r}
 - \frac{2(1-f)}{r^2} \right]\!\psi = 0,
\end{split}
\label{eq:psi_eq}
\end{equation}

where \( \psi(t,r) = \dfrac{f(r)}{r} h_1(t,r) \).

By introducing the tortoise coordinate \( dr_* = dr / f(r) \), the wave equation takes the standard Regge–Wheeler form:
\begin{equation}
\frac{\partial^2 \psi(t,r)}{\partial t^2}
 - \frac{\partial^2 \psi(t,r)}{\partial r_*^2}
 + V(r)\psi(t,r) = 0,
\label{eq:RW_wave}
\end{equation}
where the effective potential \( V(r) \) is given by
\begin{equation}
V(r) = f(r)\!\left[ \frac{\ell(\ell+1)}{r^2} - \frac{f'(r)}{r} - \frac{2}{r^2}(1 - f(r)) \right].
\label{eq:RW_potential}
\end{equation}

Equations (\ref{eq:potential}), and (\ref{eq:RW_potential}) show that the effective potentials of different fields behave differently. The effective potential governs the quasinormal modes of black holes and provides insight into particle motion, system stability, and the underlying physical properties of black holes. Figures~3 and~4 illustrate that the presence of a dark matter halo lowers the effective potential of the black hole, with the potential peak decreasing as \( r_s \) or \( \rho_s \) increase. The case \( \rho_s = 0 \) corresponds to the Schwarzschild black hole without dark matter. A lower potential reduces the energy barrier for particle motion, allowing particles to move more freely around the black hole. Consequently, the system becomes less resistant to external perturbations, making it more susceptible to changes in matter distribution and energy transfer near the black hole.

\section{WKB approach and  Padé approximants}
\label{sec:GR4}


A semi-analytic yet remarkably effective approach for computing the QNMs of black holes is based on the WKB approximation. Originally developed for quantum mechanical scattering problems, this technique was first adapted to black hole perturbations by Schutz and Will~\cite{Schutz:1985} and later refined by Iyer and Will~\cite{Iyer:1986np}. Subsequent advancements, including higher-order extensions and Padé resummation techniques~\cite{Konoplya:2003ii, Konoplya:2019hlu, Matyjasek:2017psv}, have significantly enhanced its precision across a wide variety of effective potentials.

The WKB formalism relies on the fact that the effective potential \(V(r)\) associated with black hole perturbations typically possesses a single, well-defined maximum located between the event horizon and spatial infinity. Expanding \(V(r)\) in a Taylor series around its maximum at \(r = r_0\) (or equivalently at the corresponding tortoise coordinate \(r_* = r_{*0}\)) yields
\begin{equation}
\begin{split}
V(r_*) &= V_0 
+ \frac{1}{2}V_0^{(2)}(r_* - r_{*0})^2 
+ \frac{1}{6}V_0^{(3)}(r_* - r_{*0})^3 
+ \cdots, \\\\ 
V_0^{(n)} &\equiv 
\left. \frac{d^n V}{dr_*^n} \right|_{r_{*0}} .
\end{split}
\label{eq:Vexpansion}
\end{equation}
Matching the WKB solutions across the classical turning points and imposing purely outgoing boundary conditions at both the horizon and infinity leads to the quantization condition
\begin{equation}
i\,\frac{\omega^2 - V_0}{\sqrt{-2V_0^{(2)}}}
 - \sum_{k=2}^{N} \Lambda_k(\{V_0^{(j)}\},n)
 = n + \frac{1}{2}, 
\qquad n = 0, 1, 2, \ldots,
\label{eq:WKBcondition}
\end{equation}
where \(n\) denotes the overtone number and \(\Lambda_k\) are correction terms depending on higher derivatives of the potential. Explicit expressions for \(\Lambda_k\) up to the 13th order can be found in Refs.~\cite{Iyer:1986np,Konoplya:2003ii}.  

Because the WKB expansion is asymptotic, truncating it at a finite order does not always yield a monotonic improvement in accuracy. A powerful way to enhance convergence is to treat the expansion as a formal power series in a bookkeeping parameter \(\epsilon\) and then apply a Padé resummation. Introducing
\begin{equation}
\omega^2(\epsilon) = V_0 - i\sqrt{-2V_0^{(2)}}\,\epsilon \left( n + \frac{1}{2} \right)
 + \sum_{k=2}^{N} \epsilon^k \Lambda_k,
\label{eq:omegapade1}
\end{equation}
one constructs the Padé rational approximant
\begin{equation}
P_{\tilde{m}/\tilde{n}}(\epsilon) =
\frac{\displaystyle \sum_{j=0}^{\tilde{m}} a_j \epsilon^j}
{\displaystyle 1 + \sum_{k=1}^{\tilde{n}} b_k \epsilon^k},
\qquad \tilde{m} + \tilde{n} = N,
\label{eq:Pade}
\end{equation}
whose Taylor expansion reproduces Eq.~(\ref{eq:omegapade1}) up to \(\mathcal{O}(\epsilon^{N+1})\).  
The quasinormal frequency is then approximated as
\begin{equation}
\omega = \sqrt{P_{\tilde{m}/\tilde{n}}(1)}.
\end{equation}
Balanced Padé approximants such as \([3/3]\) for \(N=6\) or \([4/4]\) for \(N=8\) generally provide the most stable and accurate results. The difference between nearby balanced approximants offers a practical estimate of the residual uncertainty. When compared with the Frobenius (Leaver) method, the sixth- or seventh-order WKB–Padé approach typically reproduces both the real and imaginary parts of the fundamental mode with relative errors below \(0.1\%\) for \(\ell \geq 1\), while maintaining reasonable accuracy even for the monopole case.

It is worth emphasizing that the WKB method is applicable only when the effective potential exhibits two distinct turning points and the wave oscillates rapidly within the barrier region. Consequently, the approximation works best for low overtone numbers (\(n < \ell\)) and for small field masses \(\mu\), where the potential maintains a single-peak structure. For massive test fields having large rest mass $\mu$, where \(V(r)\) no longer exhibits this barrier shape, one must resort to more precise methods such as the Frobenius or time-domain integration techniques.

In this work, we employ the sixth- and eighth-order WKB formalism in conjunction with Padé approximants of types \([3/3]\) split for the $6^{th}$ order WKB and \([4/4]\) split for the $8^{th}$ order WKB, which have been demonstrated in several contexts to yield optimal accuracy.

\begin{widetext}

\begin{table}[h!]
\centering
\caption{Fundamemtal ($n=0$) quasinormal mode frequencies $\omega$ for various values of $\rho_s$ and $r_s$ for scalar field perturbations. We set $l =2$ and $M=1$.}
\begin{tabular}{c c c c c c}
\hline\hline
$\rho_s$ & $r_s$ & \multicolumn{2}{c}{$6^{th}$ order WKB (Padé $\tilde{m} = 3$)} & \multicolumn{2}{c}{$8^{th}$ order WKB (Padé $\tilde{m} =4$)} \\
 & & $\mathrm{Re}(\omega)$ & $\mathrm{Im}(\omega)$ & $\mathrm{Re}(\omega)$ & $\mathrm{Im}(\omega)$ \\
\hline
0.00 & 0.0 & 0.483643 & $-0.096758i$ & 0.483643 & $-0.096758i$ \\
0.01 & 0.15 & 0.483443 & $-0.0967183i$ & 0.483444 & $-0.0967184i$  \\
0.02 & 0.25 & 0.481826 & $-0.0963861i$ & 0.481826  & $-0.0963859i$ \\
0.03 & 0.30  & 0.478995 & $-0.0958013i$ & 0.478996  & $-0.0958009i$ \\
0.04 & 0.45 & 0.463799 & $-0.0926297i$ & 0.463800 & $-0.0926295i$ \\
0.05 & 0.50 & 0.450745 & $-0.0899029i$ & 0.450746 & $-0.0899028i$ \\
0.06 & 0.65 & 0.406372  & $-0.0806318i$ & 0.406373 & $-0.0806316i$ \\
0.07 & 0.8 & 0.342448 & $-0.0674402i$ & 0.342448  & $-0.0674400i$ \\
\hline\hline
\end{tabular}
\label{Tab1}
\end{table}

\begin{table}[h!]
\centering
\caption{Fundamemtal ($n=0$) quasinormal mode frequencies $\omega$ for various values of $\rho_s$ and $r_s$ for scalar field perturbations. We set $l =3$ and $M=1$.}
\begin{tabular}{c c c c c c}
\hline\hline
$\rho_s$ & $r_s$ & \multicolumn{2}{c}{$6^{th}$ order WKB (Padé $\tilde{m} = 3$)} & \multicolumn{2}{c}{$8^{th}$ order WKB (Padé $\tilde{m} =4$)} \\
 & & $\mathrm{Re}(\omega)$ & $\mathrm{Im}(\omega)$ & $\mathrm{Re}(\omega)$ & $\mathrm{Im}(\omega)$ \\
\hline
0.00 & 0.0 & 0.675366 & $-0.0964996i$ & 0.675366 & $-0.0964996i$ \\
0.01 & 0.15 & 0.675087 & $-0.0964591i$ & 0.675087 & $-0.0964591i$  \\
0.02 & 0.25 & 0.672828 & $-0.0961278i$ & 0.672828  & $-0.0961277i$ \\
0.03 & 0.30  & 0.668874 & $-0.0955447i$ & 0.668874  & $-0.0955446i$ \\
0.04 & 0.45 & 0.647653 & $-0.0923823i$ & 0.647653 & $-0.0923823i$ \\
0.05 & 0.50 & 0.629422 & $-0.0896634i$ & 0.629422 & $-0.0896634i$ \\
0.06 & 0.65 & 0.567452 & $-0.0804192i$ & 0.567452 & $-0.0804192i$ \\
0.07 & 0.8 & 0.478181 & $-0.0672650i$ & 0.478181  & $-0.0672650i$ \\
\hline\hline
\end{tabular}
\label{Tab2}
\end{table}

\begin{table}[h!]
\centering
\caption{First overtone ($n=1$) quasinormal mode frequencies $\omega$ for various values of $\rho_s$ and $r_s$ for scalar field perturbations. We set $l =2$ and $M=1$.}
\begin{tabular}{c c c c c c}
\hline\hline
$\rho_s$ & $r_s$ & \multicolumn{2}{c}{$6^{th}$ order WKB (Padé $\tilde{m} = 3$)} & \multicolumn{2}{c}{$8^{th}$ order WKB (Padé $\tilde{m} =4$)} \\
 & & $\mathrm{Re}(\omega)$ & $\mathrm{Im}(\omega)$ & $\mathrm{Re}(\omega)$ & $\mathrm{Im}(\omega)$ \\
\hline
0.00 & 0.0 & 0.463846 & $-0.2956253i$ & 0.463847 & $-0.2956104i$ \\
0.01 & 0.15 & 0.463654 & $-0.2955010i$ & 0.463656 & $-0.2954861i$  \\
0.02 & 0.25 & 0.462109 & $-0.2944848i$ & 0.462110 & $-0.2944700i$ \\
0.03 & 0.30  & 0.459406 & $-0.2926958i$ & 0.459407 & $-0.2926811i$ \\
0.04 & 0.45 & 0.444917 & $-0.2829887i$ & 0.444918 & $-0.2829746i$ \\
0.05 & 0.50 & 0.432471 & $-0.2746425i$ & 0.432472 & $-0.2746291i$ \\
0.06 & 0.65 & 0.390163 & $-0.2462656i$ & 0.390164 & $-0.2462544i$ \\
0.07 & 0.8 & 0.329107 & $-0.2059110i$ & 0.329108  & $-0.2059024i$ \\
\hline\hline
\end{tabular}
\label{Tab3}
\end{table}

\begin{table}[h!]
\centering
\caption{First overtone ($n=1$) quasinormal mode frequencies $\omega$ for various values of $\rho_s$ and $r_s$ for scalar field perturbations. We set $l =3$ and $M=1$.}
\begin{tabular}{c c c c c c}
\hline\hline
$\rho_s$ & $r_s$ & \multicolumn{2}{c}{$6^{th}$ order WKB (Padé $\tilde{m} = 3$)} & \multicolumn{2}{c}{$8^{th}$ order WKB (Padé $\tilde{m} =4$)} \\
 & & $\mathrm{Re}(\omega)$ & $\mathrm{Im}(\omega)$ & $\mathrm{Re}(\omega)$ & $\mathrm{Im}(\omega)$ \\
\hline

0.00 & 0.0 & 0.660670 & $-0.2922874i$ & 0.660671 & $-0.2922854i$ \\
0.01 & 0.15 & 0.660397 & $-0.2921645i$ & 0.660398 & $-0.2921625i$ \\
0.02 & 0.25 & 0.658192 & $-0.2911605i$ & 0.658192 & $-0.2911585i$ \\
0.03 & 0.30  & 0.654334 & $-0.2893931i$ & 0.654334 & $-0.2893912i$ \\
0.04 & 0.45 & 0.633637 & $-0.2798060i$ & 0.633637 & $-0.2798041i$ \\
0.05 & 0.50 & 0.615857 & $-0.2715632i$ & 0.615857 & $-0.2715614i$ \\
0.06 & 0.65 & 0.555420 & $-0.2435376i$ & 0.555421 & $-0.2435360i$ \\
0.07 & 0.8 & 0.468279 & $-0.2036693i$ & 0.468279 & $-0.2036681i$ \\

\hline\hline
\end{tabular}
\label{Tab4}
\end{table}

\begin{table}[h!]
\centering
\caption{Fundamental ($n=0$) quasinormal mode frequencies $\omega$ for various values of $\rho_s$ and $r_s$ for electromagnetic field perturbations. We set $l =2$ and $M=1$.}
\begin{tabular}{c c c c c c}
\hline\hline
$\rho_s$ & $r_s$ & \multicolumn{2}{c}{$6^{th}$ order WKB (Padé $\tilde{m} = 3$)} & \multicolumn{2}{c}{$8^{th}$ order WKB (Padé $\tilde{m} =4$)} \\
 & & $\mathrm{Re}(\omega)$ & $\mathrm{Im}(\omega)$ & $\mathrm{Re}(\omega)$ & $\mathrm{Im}(\omega)$ \\
\hline

0.00 & 0.0 & 0.457594 & $-0.0950046i$ & 0.457595 & $-0.0950044i$ \\
0.01 & 0.15 & 0.457405 & $-0.0949647i$ & 0.457406 & $-0.0949645i$ \\
0.02 & 0.25 & 0.455877 & $-0.0946388i$ & 0.455878 & $-0.0946386i$ \\
0.03 & 0.30  & 0.453204 & $-0.0940654i$ & 0.453205 & $-0.0940652i$ \\
0.04 & 0.45 & 0.438865 & $-0.0909568i$ & 0.438865 & $-0.0909566i$ \\
0.05 & 0.50 & 0.426546 & $-0.0882842i$ & 0.426547 & $-0.0882841i$ \\
0.06 & 0.65 & 0.384675 & $-0.0791975i$ & 0.384675 & $-0.0791974i$ \\
0.07 & 0.8 & 0.324307 & $-0.0662612i$ & 0.324308 & $-0.0662611i$ \\

\hline\hline
\end{tabular}
\label{Tab5}
\end{table}

\begin{table}[h!]
\centering
\caption{Fundamental ($n=0$) quasinormal mode frequencies $\omega$ for various values of $\rho_s$ and $r_s$ for electromagnetic field perturbations. We set $l =3$ and $M=1$.}
\begin{tabular}{c c c c c c}
\hline\hline
$\rho_s$ & $r_s$ & \multicolumn{2}{c}{$6^{th}$ order WKB (Padé $\tilde{m} = 3$)} & \multicolumn{2}{c}{$8^{th}$ order WKB (Padé $\tilde{m} =4$)} \\
 & & $\mathrm{Re}(\omega)$ & $\mathrm{Im}(\omega)$ & $\mathrm{Re}(\omega)$ & $\mathrm{Im}(\omega)$ \\
\hline
0.00 & 0.0 & 0.656898 & $-0.0956162i$ & 0.656898 & $-0.0956162i$ \\
0.01 & 0.15 & 0.656626 & $-0.0955761i$ & 0.656627 & $-0.0955760i$ \\
0.02 & 0.25 & 0.654431 & $-0.0952480i$ & 0.654431 & $-0.0952479i$ \\
0.03 & 0.30  & 0.650590 & $-0.0946706i$ & 0.650590 & $-0.0946705i$ \\
0.04 & 0.45 & 0.629975 & $-0.0915399i$ & 0.629975 & $-0.0915398i$ \\
0.05 & 0.50 & 0.612265 & $-0.0888482i$ & 0.612265 & $-0.0888482i$ \\
0.06 & 0.65 & 0.552069 & $-0.0796967i$ & 0.552069 & $-0.0796967i$ \\
0.07 & 0.8 & 0.465319 & $-0.0666710i$ & 0.465319 & $-0.0666709i$ \\
\hline\hline
\end{tabular}
\label{Tab6}
\end{table}

\begin{table}[h!]
\centering
\caption{First overtone ($n=1$) quasinormal mode frequencies $\omega$ for various values of $\rho_s$ and $r_s$ for electromagnetic field perturbations. We set $l =2$ and $M=1$.}
\begin{tabular}{c c c c c c}
\hline\hline
$\rho_s$ & $r_s$ & \multicolumn{2}{c}{$6^{th}$ order WKB (Padé $\tilde{m} = 3$)} & \multicolumn{2}{c}{$8^{th}$ order WKB (Padé $\tilde{m} =4$)} \\
 & & $\mathrm{Re}(\omega)$ & $\mathrm{Im}(\omega)$ & $\mathrm{Re}(\omega)$ & $\mathrm{Im}(\omega)$ \\
\hline
0.00 & 0.0 & 0.436532 & $-0.2907260i$ & 0.436534 & $-0.2907193i$ \\
0.01 & 0.15 & 0.436352 & $-0.2906038i$ & 0.436354 & $-0.2905971i$ \\
0.02 & 0.25 & 0.434901 & $-0.2896055i$ & 0.434903 & $-0.2895988i$ \\
0.03 & 0.30  & 0.432366 & $-0.2878483i$ & 0.432366 & $-0.2878416i$ \\
0.04 & 0.45 & 0.418775 & $-0.2783177i$ & 0.418777 & $-0.2783110i$ \\
0.05 & 0.50 & 0.407103 & $-0.2701235i$ & 0.407105 & $-0.2701169i$ \\
0.06 & 0.65 & 0.367428 & $-0.2422635i$ & 0.367430 & $-0.2422572i$ \\
0.07 & 0.8 & 0.310111 & $-0.2026235i$ & 0.310113 & $-0.2026177i$ \\
\hline\hline
\end{tabular}
\label{Tab7}
\end{table}

\begin{table}[h!]
\centering
\caption{First overtone ($n=1$) quasinormal mode frequencies $\omega$ for various values of $\rho_s$ and $r_s$ for electromagnetic field perturbations. We set $l =3$ and $M=1$.}
\begin{tabular}{c c c c c c}
\hline\hline
$\rho_s$ & $r_s$ & \multicolumn{2}{c}{$6^{th}$ order WKB (Padé $\tilde{m} = 3$)} & \multicolumn{2}{c}{$8^{th}$ order WKB (Padé $\tilde{m} =4$)} \\
 & & $\mathrm{Re}(\omega)$ & $\mathrm{Im}(\omega)$ & $\mathrm{Re}(\omega)$ & $\mathrm{Im}(\omega)$ \\
\hline
0.00 & 0.0 & 0.641736 & $-0.2897303i$ & 0.641736 & $-0.2897292i$ \\
0.01 & 0.15 & 0.641471 & $-0.2896086i$ & 0.641471 & $-0.2896074i$ \\
0.02 & 0.25 & 0.639331 & $-0.2886138i$ & 0.639331 & $-0.2886127i$ \\
0.03 & 0.30  & 0.635587 & $-0.2868630i$ & 0.635587 & $-0.2868619i$ \\
0.04 & 0.45 & 0.615513 & $-0.2773676i$ & 0.615513 & $-0.2773665i$ \\
0.05 & 0.50 & 0.598269 & $-0.2692038i$ & 0.598269 & $-0.2692027i$ \\
0.06 & 0.65 & 0.539654 & $-0.2414470i$ & 0.539654 & $-0.2414460i$ \\
0.07 & 0.8 & 0.455100 & $-0.2019506i$ & 0.455101 & $-0.2019497i$ \\
\hline\hline
\end{tabular}
\label{Tab8}
\end{table}

\begin{table}[h!]
\centering
\caption{Fundamental ($n=0$) quasinormal mode frequencies $\omega$ for various values of $\rho_s$ and $r_s$ for gravitational field perturbations. We set $l =2$ and $M=1$.}
\begin{tabular}{c c c c c c}
\hline\hline
$\rho_s$ & $r_s$ & \multicolumn{2}{c}{$6^{th}$ order WKB (Padé $\tilde{m} = 3$)} & \multicolumn{2}{c}{$8^{th}$ order WKB (Padé $\tilde{m} =4$)} \\
 & & $\mathrm{Re}(\omega)$ & $\mathrm{Im}(\omega)$ & $\mathrm{Re}(\omega)$ & $\mathrm{Im}(\omega)$ \\
\hline

0.00 & 0.0 & 0.373619 & $-0.0889327i$ & 0.373669 & $-0.0889722i$ \\
0.01 & 0.15 & 0.373465 & $-0.0888954i$ & 0.373514 & $-0.0889348i$ \\
0.02 & 0.25 & 0.372218 & $-0.0885904i$ & 0.372267 & $-0.0886297i$ \\
0.03 & 0.30  & 0.370035 & $-0.0880538i$ & 0.370083 & $-0.0880928i$ \\
0.04 & 0.45 & 0.358326 & $-0.0851451i$ & 0.358370 & $-0.0851827i$ \\
0.05 & 0.50 & 0.348267 & $-0.0826445i$ & 0.348308 & $-0.0826808i$ \\
0.06 & 0.65 & 0.314076 & $-0.0741432i$ & 0.314107 & $-0.0741748i$ \\
0.07 & 0.8 & 0.264783 & $-0.0620395i$ & 0.264801 & $-0.0620638i$ \\

\hline\hline
\end{tabular}
\label{Tab9}
\end{table}

\begin{table}[h!]
\centering
\caption{Fundamental ($n=0$) quasinormal mode frequencies $\omega$ for various values of $\rho_s$ and $r_s$ for gravitational field perturbations. We set $l =3$ and $M=1$.}
\begin{tabular}{c c c c c c}
\hline\hline
$\rho_s$ & $r_s$ & \multicolumn{2}{c}{$6^{th}$ order WKB (Padé $\tilde{m} = 3$)} & \multicolumn{2}{c}{$8^{th}$ order WKB (Padé $\tilde{m} =4$)} \\
 & & $\mathrm{Re}(\omega)$ & $\mathrm{Im}(\omega)$ & $\mathrm{Re}(\omega)$ & $\mathrm{Im}(\omega)$ \\
\hline
0.00 & 0.0 & 0.599443 & $-0.0927029i$ & 0.599443 & $-0.0927029i$ \\
0.01 & 0.15 & 0.599195 & $-0.0926639i$ & 0.599195 & $-0.0926640i$ \\
0.02 & 0.25 & 0.597192 & $-0.0923460i$ & 0.597192 & $-0.0923460i$ \\
0.03 & 0.30  & 0.593686 & $-0.0917866i$ & 0.593686 & $-0.0917866i$ \\
0.04 & 0.45 & 0.574873 & $-0.0887538i$ & 0.574873 & $-0.0887539i$ \\
0.05 & 0.50 & 0.558712 & $-0.0861465i$ & 0.558712 & $-0.0861465i$ \\
0.06 & 0.65 & 0.503777 & $-0.0772817i$ & 0.503777 & $-0.0772817i$ \\
0.07 & 0.8 & 0.424612 & $-0.0646606i$ & 0.424612 & $-0.0646606i$ \\
\hline\hline
\end{tabular}
\label{Tab10}
\end{table}

\begin{table}[h!]
\centering
\caption{First overtone ($n=1$) quasinormal mode frequencies $\omega$ for various values of $\rho_s$ and $r_s$ for gravitational field perturbations. We set $l =2$ and $M=1$.}
\begin{tabular}{c c c c c c}
\hline\hline
$\rho_s$ & $r_s$ & \multicolumn{2}{c}{$6^{th}$ order WKB (Padé $\tilde{m} = 3$)} & \multicolumn{2}{c}{$8^{th}$ order WKB (Padé $\tilde{m} =4$)} \\
 & & $\mathrm{Re}(\omega)$ & $\mathrm{Im}(\omega)$ & $\mathrm{Re}(\omega)$ & $\mathrm{Im}(\omega)$ \\
\hline
0.00 & 0.0 & 0.346007 & $-0.2735657i$ & 0.346002 & $-0.2735551i$ \\
0.01 & 0.15 & 0.345865 & $-0.2734508i$ & 0.345860 & $-0.2734401i$ \\
0.02 & 0.25 & 0.344718 & $-0.2725114i$ & 0.344713 & $-0.2725006i$ \\
0.03 & 0.30  & 0.342715 & $-0.2708581i$ & 0.342710 & $-0.2708471i$ \\
0.04 & 0.45 & 0.332001 & $-0.2618917i$ & 0.331995 & $-0.2618790i$ \\
0.05 & 0.50 & 0.322799 & $-0.2541830i$ & 0.322793 & $-0.2541686i$ \\
0.06 & 0.65 & 0.291516 & $-0.2279753i$ & 0.291509 & $-0.2279541i$ \\
0.07 & 0.8 & 0.246252 & $-0.1906849i$ & 0.246245 & $-0.1906537i$ \\
\hline\hline
\end{tabular}
\label{Tab11}
\end{table}

\begin{table}[h!]
\centering
\caption{First overtone ($n=1$) quasinormal mode frequencies $\omega$ for various values of $\rho_s$ and $r_s$ for gravitational field perturbations. We set $l =3$ and $M=1$.}
\begin{tabular}{c c c c c c}
\hline\hline
$\rho_s$ & $r_s$ & \multicolumn{2}{c}{$6^{th}$ order WKB (Padé $\tilde{m} = 3$)} & \multicolumn{2}{c}{$8^{th}$ order WKB (Padé $\tilde{m} =4$)} \\
 & & $\mathrm{Re}(\omega)$ & $\mathrm{Im}(\omega)$ & $\mathrm{Re}(\omega)$ & $\mathrm{Im}(\omega)$ \\
\hline

0.00 & 0.0 & 0.582640 & $-0.2812888i$ & 0.582644 & $-0.2812984i$ \\
0.01 & 0.15 & 0.582399 & $-0.2811707i$ & 0.582404 & $-0.2811802i$ \\
0.02 & 0.25 & 0.580457 & $-0.2802054i$ & 0.580461 & $-0.2802149i$ \\
0.03 & 0.30  & 0.577059 & $-0.2785066i$ & 0.577064 & $-0.2785160i$ \\
0.04 & 0.45 & 0.558845 & $-0.2692955i$ & 0.558849 & $-0.2693043i$ \\
0.05 & 0.50 & 0.543198 & $-0.2613763i$ & 0.543202 & $-0.2613845i$ \\
0.06 & 0.65 & 0.490013 & $-0.2344512i$ & 0.490016 & $-0.2344576i$ \\
0.07 & 0.8 & 0.413278 & $-0.1961286i$ & 0.413281 & $-0.1961329i$ \\

\hline\hline
\end{tabular}
\label{Tab12}
\end{table}

\end{widetext}

As seen from Tables (\ref{Tab1}) and (\ref{Tab2}), which present the fundamental ($n = 0$) quasinormal mode frequencies of scalar field perturbations for $\ell = 2$ and $\ell = 3$, a clear and consistent trend is observed. As the DM-halo parameters $\rho_s$ and $r_s$ increase, the real part of the frequency $Re(\omega)$ decreases, indicating a reduction in the characteristic oscillation frequency of the resulting ringdown waveform. Likewise, the magnitude of the imaginary part $|Im(\omega)|$, which quantifies the damping rate of the perturbations, also decreases. This behaviour implies a longer relaxation timescale and the presence of longer-lived quasinormal modes in the black hole--halo system. Physically, this means that the rate of energy loss during the evolution of the perturbations is reduced when the surrounding halo becomes denser or more extended, allowing the system to maintain the perturbative state for a longer duration. Additionally, noticeable deviations from the standard Schwarzschild values begin to appear when the halo parameters reach approximately $\rho_s \approx 0.03$ and $r_s \approx 0.3$, while the strongest modifications occur near $\rho_s = 0.07$ and $r_s = 0.8$, which represent the upper range of parameters considered in this work.
A similar behaviour is observed for the first overtone ($n = 1$) quasinormal mode frequencies in tables (\ref{Tab3}) and (\ref{Tab4}). As the halo parameters $\rho_s$ and $r_s$ increase, the real part of the overtone frequency shows a steady decrease, signalling a reduction in the oscillation rate of the higher mode. Simultaneously, the magnitude of the imaginary component $|Im(\omega)|$ also decreases, indicating weaker damping and a longer lifetime for the overtone in the presence of a denser or more extended DM halo. This demonstrates that the influence of the halo is not limited to the fundamental mode but extends consistently across the entire quasinormal spectrum.

In Tables (\ref{Tab5}) to (\ref{Tab8}), we present the QNM frequencies for the 
electromagnetic perturbations, including both the fundamental mode and the first overtone for the multipole numbers $l = 2$ and $l = 3$. 
The overall behaviour closely mirrors the trends observed in the scalar field case. As the DM-halo parameters $\rho_s$ and $r_s$ 
increase, the real part of the frequency $Re(\omega)$ exhibits a pronounced decrease, leading to a lower oscillation frequency of the 
resulting ringdown signal. At the same time, the magnitude of the imaginary part $|Im(\omega)|$ also decreases, indicating a slower 
damping rate and correspondingly longer--lived electromagnetic QNMs. These deviations from the Schwarzschild values become 
progressively more significant for higher values of $\rho_s$ and $r_s$, and each pair of halo parameters yields a distinct set of QNM 
frequencies. This demonstrates that the presence of the Dehnen--type dark--matter halo imprints a clear and detectable modification on the 
electromagnetic perturbation spectrum as well.

Finally, we come to the gravitational field perturbation case, which is 
the most important one since gravitational perturbations arise directly 
from black hole mergers and are observed in the gravitational--wave 
spectrum. The trends in tables (\ref{Tab9}) to (\ref{Tab12}) displayed in the corresponding QNM tables closely follow those seen in the scalar and electromagnetic sectors. As the halo 
parameters $\rho_s$ and $r_s$ increase, the real part of the frequency $Re(\omega)$ decreases, indicating a reduction in the characteristic 
oscillation frequency of the gravitational ringdown signal. Likewise, the magnitude of the imaginary part $|Im(\omega)|$ decreases, implying weaker damping and correspondingly longer--lived gravitational QNMs. These deviations become more pronounced for larger halo strengths, confirming that the presence of a DM halo leaves a detectable imprint on the gravitational QNM spectrum, which is precisely the sector most relevant for current and future gravitational wave observations.

Across all the tables, we have systematically demonstrated the behaviour of the QNM frequencies for the scalar, electromagnetic, and gravitational perturbations. For consistency, our results were cross-checked using the eighth--order WKB approximation with the $[4/4]$ Padé splitting, which agrees remarkably well with the 
sixth--order WKB results presented throughout the analysis. As the two parameters of the BH--DM halo metric, $\rho_s$ and $r_s$, are varied, 
the oscillation and damping characteristics of the system change in a coordinated manner. This reflects the synergistic influence of the 
dark--matter halo density and the black hole's gravitational field, both of which are deeply intertwined in determining the dynamical response of the spacetime. The resulting modifications to the quasinormal spectra 
offer potentially detectable signatures for future gravitational--wave observations. Such findings provide valuable insights into the physical mechanisms governing the interaction between dark matter and black holes, and studies of this kind may ultimately contribute to uncovering new aspects of dark matter as a candidate beyond the Standard Model of 
particle physics.


\section{Particle motion and Shadow radius}
\label{sec:GR5}

Black hole shadows open a direct window onto the spacetime geometry in the strong-field regime of gravity, the dark “silhouette” is set by unstable photon orbits near the photon sphere and, in general relativity, its angular size depends mainly on the black hole’s mass-to-distance ratio and only weakly on spin or viewing angle.

In order to examine the particle trajectory around this BH-DM halo system we investigate the motion of test particles and photons in a curved spacetime by the geodesic equations, which can be derived from the Euler--Lagrange equation
\begin{equation}
\frac{d}{d\tau}\!\left(\frac{\partial \mathcal{L}}{\partial \dot{x}^{\mu}}\right)
-
\frac{\partial \mathcal{L}}{\partial x^{\mu}} = 0,
\label{EL_equation}
\end{equation}
where $\tau$ is an affine parameter along the worldline. For photon motion near a
Schwarzschild-like black hole surrounded by a Dehnen-type DM halo,
we begin with the Lagrangian
\begin{equation}
\mathcal{L} = \frac{1}{2} g_{\mu\nu}\dot{x}^{\mu}\dot{x}^{\nu}.
\end{equation}
For the line element (\ref{eq:metric-dehnen}), this becomes
\begin{equation}
\mathcal{L}
= \frac{1}{2}
\left[
-f(r)\dot{t}^{2}
+\frac{\dot{r}^{2}}{f(r)}
+r^{2}\dot{\theta}^{2}
+r^{2}\sin^{2}\theta\,\dot{\phi}^{2}
\right].
\end{equation}

Because the spacetime is static and spherically symmetric, the corresponding Killing
vectors lead to the conserved quantities
\begin{equation}
E = f(r)\dot{t}, \qquad
L = r^{2}\sin^{2}\theta\,\dot{\phi}.
\end{equation}
Photons satisfy the null condition $\mathcal{L}=0$, and by symmetry we restrict the motion 
to the equatorial plane $\theta=\pi/2$. After rescaling the affine parameter 
$\tau \rightarrow \tau/L$, the equations of motion reduce to
\begin{equation}
\dot{t}=\frac{1}{b f(r)}, \qquad
\dot{\phi}=\pm\frac{1}{r^{2}}, \qquad
\dot{r}^{2}=\frac{1}{b^{2}} - V_{\rm eff}(r),
\end{equation}
where $b=L/E$ is the impact parameter. The effective potential governing the radial 
motion is
\begin{equation}
V_{\rm eff}(r) = \frac{f(r)}{r^{2}}.
\end{equation}

Using the relation between $r$ and $\phi$, we obtain the trajectory equation
\begin{equation}
\left(\frac{dr}{d\phi}\right)^{2}
= r^{4}\left[\frac{1}{b^{2}} - V_{\rm eff}(r) \right].
\end{equation}

The motion of photons is therefore highly sensitive to the impact parameter and the
structure of the effective potential. Depending on its value of $b$, a photon may escape to
infinity, fall into the black hole, or asymptotically approach an unstable circular orbit—the
photon sphere. This unstable orbit determines the boundary of the black hole shadow. The
photon sphere is defined by the conditions
\begin{equation}
V_{\rm eff}(r_{\rm ph}) = \frac{1}{b_{\rm ph}^{2}}, 
\qquad
V'_{\rm eff}(r_{\rm ph})=0,
\end{equation}
which yield the critical impact parameter
\begin{equation}
b_{\rm ph} = \frac{r_{\rm ph}}{\sqrt{f(r_{\rm ph})}}.
\end{equation}

We examined the behaviour of the photon sphere radius $r_{\text{ph}}$ and the corresponding critical impact parameter $b_{\text{ph}}$, both of which depend sensitively on the dark--matter 
halo parameters $\rho_s$ and $r_s$. Variations in $b_{\text{ph}}$ directly translate into changes in the apparent size of the black hole 
shadow as seen by a distant observer. Consequently, any modification induced by the surrounding dark--matter distribution provides an opportunity to place meaningful constraints on the allowed ranges of $\rho_s$ and $r_s$ using current and future observations of black hole shadows.

The fundamental equations governing the radius of a black hole shadow 
have been well established in the literature \cite{Synge:1966, Claudel:2000yi} and have 
been applied extensively in a wide range of studies (see, for instance, 
\cite{Virbhadra:2002ju, Perlick:2021aok, Bisnovatyi-Kogan:2017kii, Konoplya:2019sns, Gan:2021xdl,Luo:2025xjb} and references therein). For a 
static, spherically symmetric spacetime, the radius of the circular 
photon orbit $r_{\text{ph}}$ is determined by the condition \cite{EventHorizonTelescope:2020qrl}
\begin{equation}
r f'(r) - 2 f(r) = 0,
\label{eq:rph_condition}
\end{equation}
which identifies the location of the unstable null geodesic. The 
corresponding shadow radius $R_{\text{sh}}$ observed by a distant 
observer is then given by
\begin{equation}
R_{\text{sh}} = \sqrt{ \frac{r_{\text{ph}}^{\,2}}{ f(r_{\text{ph}}) } },
\label{eq:Rsh_definition}
\end{equation}
providing a direct link between the spacetime geometry and the apparent size of the black hole shadow. These relations form the basis for connecting theoretical models with observational constraints from black hole imaging experiments.

\begin{figure}[h!]
    \centering
    \includegraphics[width=0.49\textwidth]{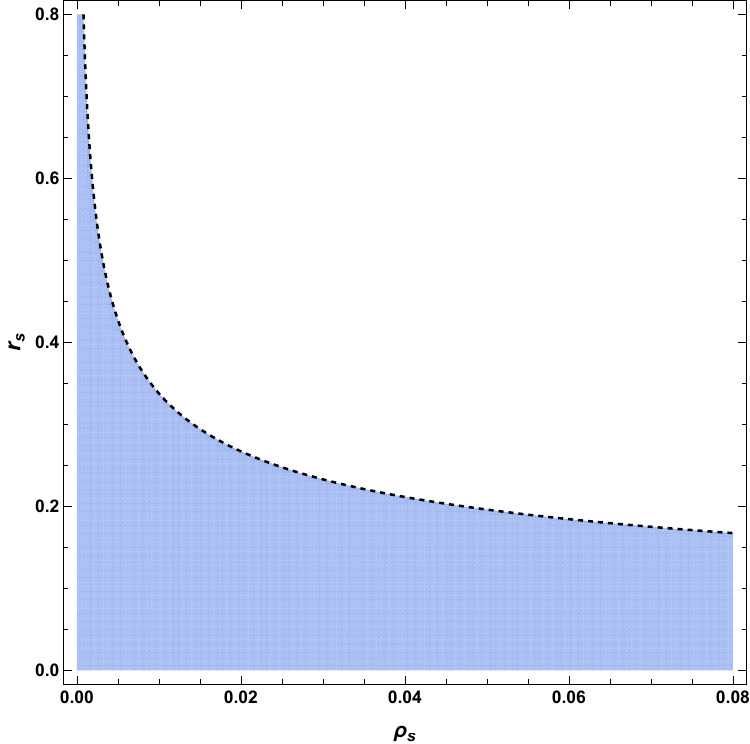}
    \caption{Allowed parameter space for the halo density $\rho_s$ and scale radius 
$r_s$ obtained from the Sgr A$^*$ shadow measurement. The dashed contour denotes the maximum shadow radius of $5.22$, and all points lying below this boundary in the blue shaded region satisfy the observational constraint.}
\label{FIG7}
\end{figure}

We can use the recent Event Horizon Telescope (EHT) observations of 
black hole shadows to constrain the parameters $\rho_s$ and $r_s$ of the 
Schwarzschild--like BH--DM halo spacetime. From the measured shadow size 
of the Sgr A$^*$ black hole \cite{EventHorizonTelescope:2019kwo}, one obtains the $1\sigma$ 
bound \cite{Vagnozzi:2022moj}
\begin{equation}
4.55\,M \;\lesssim\; R_{\text{sh}} \;\lesssim\; 5.22\,M .
\label{eq:EHT_bound}
\end{equation}
Motivated by this, in Fig.(\ref{FIG7}) we present the corresponding constrained parameter space in the $(\rho_s, r_s)$ plane for a Schwarzschild--like black hole surrounded by a Dehnen--type $(1,4,2)$ dark--matter halo. The 
black dashed contour in Fig.(\ref{FIG7}) represents the curve $R_{\text{sh}} = 5.22\,M$, and the region enclosed by this contour together with the coordinate axes corresponds to shadow radii smaller than this observational upper limit. This region, therefore, defines the 
allowed range of the halo parameters $\rho_s$ and $r_s$. All parameter values used in the subsequent analysis lie within this observationally permitted domain.


\section{Grey-body Factors}
\label{sec:GR6}


Grey-body factors describe the fraction of Hawking radiation that is able to
penetrate the effective potential barrier surrounding the black hole rather than
being reflected back toward the event horizon. To evaluate these factors, we use
Hawking's semiclassical radiation formula supplemented with the grey-body
modification, allowing us to compute the radiation flux that reaches a distant
observer. This approach remains valid even during the late stages of black hole
evaporation and for the modified geometry specified by the metric function
(Eq.~(2)).

It is well established that the contribution of gravitons to the Hawking flux is
extremely small---in the Schwarzschild case, less than \(2\%\) of the total
emission~\cite{Page:1976df}. Consequently, grey-body factors computed for test
fields provide an accurate characterization of the radiation spectrum. In fact,
these factors often play a more significant role than the Hawking temperature in
determining the emitted flux~\cite{Konoplya:2019ppy}.

To compute the grey-body factors, we analyze the wave equation under scattering
boundary conditions that allow an incoming wave from spatial infinity. Owing to
the symmetry of the scattering process, this is equivalent to considering a wave
incident from the horizon. The boundary conditions for the radial field
\(\Psi(r_*)\) are
\begin{equation}
\Psi(r_*) =
\begin{cases}
e^{-i\omega r_*} + R e^{i\omega r_*}, & r_* \to +\infty, \\[6pt]
T e^{-i\omega r_*}, & r_* \to -\infty,
\end{cases}
\end{equation}
where \(R\) and \(T\) denote the reflection and transmission amplitudes,
respectively.

Because the effective potential forms a single barrier and decreases
monotonically in both asymptotic regions, the WKB approximation can be reliably
applied to compute the scattering amplitudes~\cite{Konoplya:2003ii}. For real
\(\omega^2\), the first-order WKB approximation yields real coefficients
satisfying
\begin{equation}
|T|^{2} + |R|^{2} = 1.
\end{equation}
Thus, the grey-body factor for a given multipole number \(\ell\) is
\begin{equation}
|A_\ell|^{2} = |T_\ell|^2 = 1 - |R_\ell|^{2}.
\end{equation}

For accurate results, we use the higher-order WKB expansion
\cite{Konoplya:2019hlu, Matyjasek:2017psv}. At very low frequencies, the
WKB approximation becomes unreliable because nearly the entire wave is reflected;
however, the contribution of this regime to the total luminosity is negligible,
and we smoothly extrapolate the WKB expression to small \(\omega\).

Following Refs.~\cite{Konoplya:2003ii, Iyer:1986np, Konoplya:2024lir, Dubinsky:2024vbn, Pathrikar:2025gzu}, the reflection amplitude can be expressed as
\begin{equation}
\Gamma_\ell(\Omega)
=
\left( 1 + e^{-2 i \pi K} \right)^{-1/2},
\end{equation}
where \(K\) is determined by
\begin{equation}
K - i \frac{\omega^{2} - V_{\text{max}}}
{\sqrt{-2 V''_{\text{max}}}}
- \sum_{i=2}^{6} \Lambda_i(K) = 0.
\end{equation}
Here, \(V_{\text{max}}\) and \(V''_{\text{max}}\) are the value and second
derivative of the effective potential at its maximum, and \(\Lambda_i(K)\) denote
the higher-order WKB correction terms.

The WKB expansion is asymptotic rather than convergent, and its optimal accuracy
typically occurs at a particular order that depends sensitively on the structure
of the effective potential. Moreover, the WKB method may fail even for large
\(\ell\) when the potential deviates from the standard centrifugal barrier
\(f(r)\ell(\ell+1)/r^{2}\). Such breakdowns arise, for example, in modified
gravity theories with higher-curvature corrections or in cases where the
perturbations become unstable~\cite{Konoplya:2020bxa, Takahashi:2011du, Gleiser:2005ra, Konoplya:2017lhs, Konoplya:2017ymp}. A systematic
discussion of situations where the WKB method fails or becomes incomplete in the
eikonal limit can be found in~\cite{Bolokhov:2023dxq, Konoplya:2022gjp}.

\begin{figure*}[t]
\centering

\resizebox{0.45\textwidth}{!}{
    \includegraphics{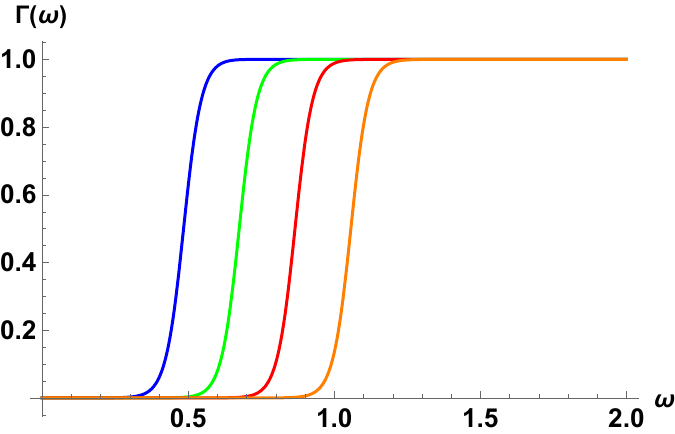}
}
\hfill
\resizebox{0.45\textwidth}{!}{
    \includegraphics{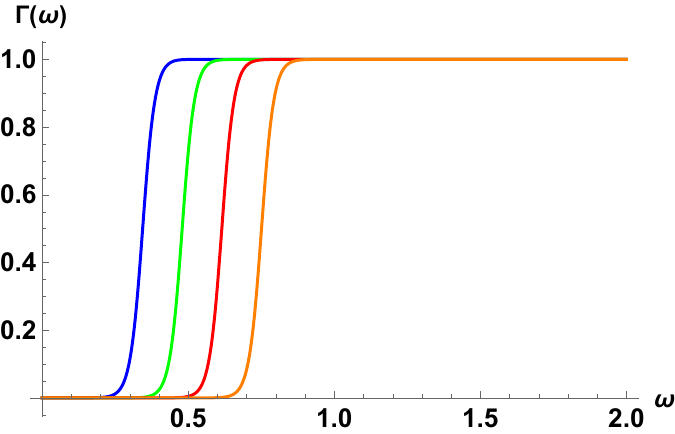}
}

\caption{Greybody factors $|\Gamma_{\ell}(\omega)|^2$ for scalar field perturbations 
in the Schwarzschild-like black hole surrounded by a Dehnen-type dark-matter 
halo. The left panel corresponds to the parameters $\rho_s = 0.02$ and 
$r_s = 0.25$, while the right panel shows the case $\rho_s = 0.07$ and 
$r_s = 0.8$ with $l = 2$ (blue), $l =3$ (green), $l=4$ (red), $l=5$ (orange).}

\label{GBF1}

\vspace{0.4cm}

\resizebox{0.45\textwidth}{!}{
    \includegraphics{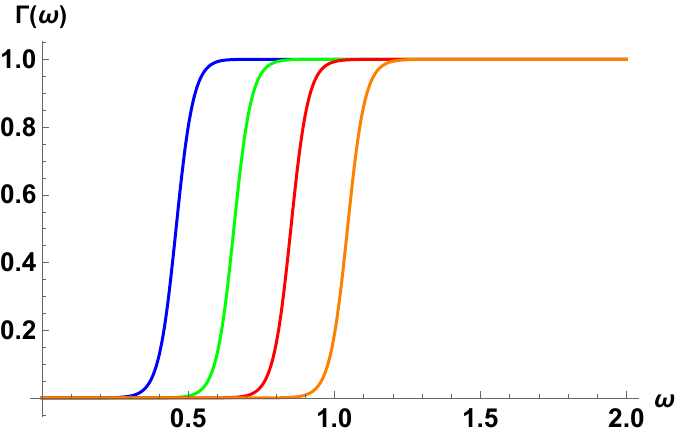}}
\hfill
\resizebox{0.45\textwidth}{!}{
    \includegraphics{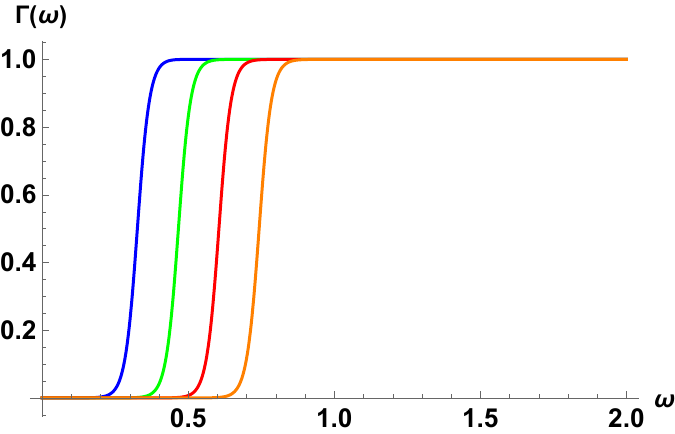}
}

\caption{Greybody factors $|\Gamma_{\ell}(\omega)|^2$ for electromagnetic field perturbations in the Schwarzschild-like black hole surrounded by a Dehnen-type dark-matter 
halo. The left panel corresponds to the parameters $\rho_s = 0.02$ and 
$r_s = 0.25$, while the right panel shows the case $\rho_s = 0.07$ and 
$r_s = 0.8$ with $l = 2$ (blue), $l =3$ (green), $l=4$ (red), $l=5$ (orange).}

\label{GBF2}

\vspace{0.4cm}

\resizebox{0.45\textwidth}{!}{
    \includegraphics{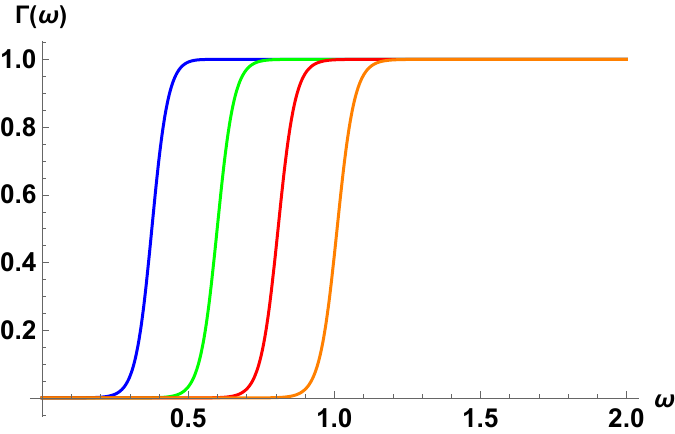}
}
\hfill
\resizebox{0.45\textwidth}{!}{
    \includegraphics{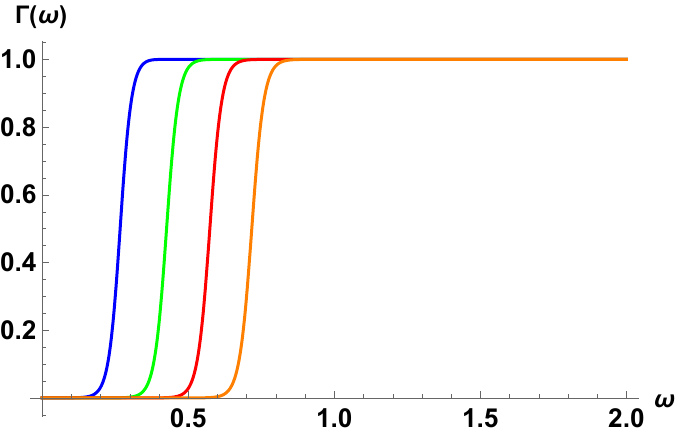}
}

\caption{Greybody factors $|\Gamma_{\ell}(\omega)|^2$ for gravitational field perturbations in the Schwarzschild-like black hole surrounded by a Dehnen-type dark-matter 
halo. The left panel corresponds to the parameters $\rho_s = 0.02$ and 
$r_s = 0.25$, while the right panel shows the case $\rho_s = 0.07$ and 
$r_s = 0.8$ with $l = 2$ (blue), $l =3$ (green), $l=4$ (red), $l=5$ (orange).
}
\label{fig:GBF3}
\end{figure*}

As we can see from the figures (\ref{GBF1}), (\ref{GBF2}), and (\ref{fig:GBF3}) the grey-body factor plots clearly show that the surrounding Dehnen-type DM halo significantly influences how radiation interacts with the black hole's effective potential barrier. As the halo density $\rho_s$ and $r_s$ increase, the barrier becomes weaker, causing the transmission curves to shift toward lower frequencies. This means that the transmission coefficients are enhanced, with radiation beginning to pass through the barrier at smaller values of $\omega$. It is also notable that this enhancement occurs universally across scalar, electromagnetic, and gravitational perturbations, and is fully consistent with the corresponding effective potentials derived for these fields, all of which decrease in height in the presence of a denser DM halo.

\clearpage
\section{Discussion and Conclusions}
\label{sec:GR7}

In this work, we have carried out a comprehensive analysis of the 
fundamental and first--overtone quasinormal mode frequencies of a 
Schwarzschild--like black hole surrounded by a Dehnen--type dark--matter 
halo characterized by the $(1,4,2)$ profile. We derived the 
corresponding wave equations and effective potentials for scalar, 
electromagnetic, and gravitational perturbations, and employed the 
sixth--order WKB approximation to compute the QNM spectra for various 
multipole numbers and halo parameters $\rho_s$ and $r_s$. Our results 
show that increasing the values of these parameters systematically 
decreases the real part of the QNM frequency, indicating a reduction in 
the oscillation frequency, and simultaneously reduces the magnitude of 
the imaginary part, implying longer-lived perturbations. This leads to 
a consistent pattern of increasingly long-lived QNMs as 
the influence of the surrounding halo becomes stronger.

To ensure accuracy, we further applied Padé resummation techniques, 
using the $[3/3]$ Padé approximant for the sixth--order WKB method, and 
verified our findings with the eighth--order WKB approximation employing 
a $[4/4]$ Padé split. The excellent agreement between these two 
independent computations provides strong confidence in the robustness of 
our results.

In addition to the perturbative analysis, we examined particle dynamics 
and photon geodesics in this spacetime. Using the standard formalism for 
black hole shadows, we constrained the halo parameters $\rho_s$ and 
$r_s$ by incorporating the Event Horizon Telescope (EHT) measurement of 
the M87* shadow radius. This allowed us to identify and visualize the 
region in parameter space consistent with current observational bounds.

Finally, we investigated the greybody factors associated with Hawking 
radiation, which encode the transmission probabilities through the 
effective potential barriers. Our analysis shows that as the halo 
parameters $\rho_s$ and $r_s$ increase, the transmission probability is 
consistently enhanced across all perturbation spins and multipole 
numbers. This behaviour is fully consistent with the structure of the 
effective potentials, whose peak heights decrease in the presence of a 
stronger halo, thereby enabling radiation to traverse the barrier more 
efficiently. Together, these results highlight that the dark--matter 
environment exerts a measurable influence on both the dynamical and 
radiative properties of black holes, offering potential observational 
signatures for future gravitational-wave and black hole imaging experiments.

This work opens several promising avenues for future investigation. A 
natural extension would be to examine Dirac/neutrino field perturbations in the 
same dark--matter environment, allowing us to determine how fermionic 
fields modify the QNM spectrum and greybody factors, and to explore 
whether their interaction with the halo leads to qualitatively new 
features. It would also be valuable to analyze the polar (even--parity) 
sector by performing a full time--domain evolution, which would help 
establish the dynamical stability of the system beyond the axial 
perturbations considered here. Another interesting direction is to test 
the correspondence between QNMs and greybody factors more 
thoroughly, particularly in the context of Hawking radiation of Dirac 
particles, where one could compute emission rates and the resulting 
spectra. Such studies would deepen our understanding of how dark matter 
influences both the oscillatory and radiative properties of black holes, 
and may provide further observationally relevant signatures.


\section*{Acknowledgments}

The author expresses his gratitude to Pankaj S. Joshi and Parth Bambhaniya for giving the opportunity to visit ICSC, Ahmedabad University. 


\end{document}